\documentstyle[12pt,epsfig]{article}
\newcommand{\ba}{\begin{array}}
\newcommand{\ea}{\end{array}}
\newcommand{\bd}{\begin{displaymath}}
\newcommand{\ed}{\end{displaymath}}
\newcommand{\be}{\begin{equation}}
\newcommand{\ee}{\end{equation}}
\newcommand{\bea}{\begin{eqnarray}}
\newcommand{\eea}{\end{eqnarray}}
\newcommand{\Dir}{\kern -6.4pt\Big{/}}
\newcommand{\Dirin}{\kern -10.4pt\Big{/}\kern 4.4pt}
\newcommand{\DDir}{\kern -10.6pt\Big{/}}
\newcommand{\DGir}{\kern -6.0pt\Big{/}}
\begin{document}
\
\def\bra{\langle}
\def\ket{\rangle}

\def\a{\alpha}
\def\as {\alpha_s}
\def\b{\beta}
\def\d{\delta}
\def\e{\epsilon}
\def\ve{\varepsilon}
\def\l{\lambda}
\def\m{\mu}
\def\n{\nu}
\def\G{\Gamma}
\def\D{\Delta}
\def\L{\Lambda}
\def\s{\sigma}
\def\p{\pi}

\def\etal{ {\em et al.}}
\def\mzs {M_Z^2}
\def\mws {M_W^2}
\def\q2 {q^2}
\def\sz {\sin^2\theta_W}
\def\cz {\cos^2\theta_W}
\def\lp{\lambda^{\prime}}
\def\lps{\lambda^{\prime *}}
\def\lpp{\lambda^{\prime\prime}}
\def\lpps{\lambda^{\prime\prime * }}

\def\bapp{b_1^{\prime\prime}}
\def\bbpp{b_2^{\prime\prime}}
\def\bcp{b_3^{\prime}}
\def\bdp{b_4^{\prime}}
\def\t {\times }
\def\slash {\!\!\!\!\!\!/}
\def\photino {\tilde\gamma}
\def\sel {\tilde{e}}
 \def\N10{\widetilde \chi_1^0}
                         \def\C1p{\widetilde \chi_1^+}
                         \def\C1m{\widetilde \chi_1^-}
                         \def\C1pm{\widetilde \chi_1^\pm}
 \def\Ntwo{\widetilde \chi_2^0}
                         \def\Ctwo{\widetilde \chi_2^\pm}
\def\lslep {{\tilde e}_L}
\def\rslep {{\tilde e}_R}
\def\sneu {\tilde \nu}
\def\msneu {M_\tilde \nu}
\def\mrslep {m_{\rslep}}
\def\mlslep {m_{\lslep}}
\def\mneu {m_{\neu}}
\def\mpT{p_T \hspace{-1em}/\;\:}
\def\mET{E_T \hspace{-1.1em}/\;\:}
\def\mE{E \hspace{-.7em}/\;\:}
\def\go{\rightarrow}
\def\beq{\begin{eqnarray}}
\def\Rp{R\!\!\!\!/}
\def\wrp {{\cal W}_{R\!\!\!\!/}}
\def\enq{\end{eqnarray}}
\def\goes{\longrightarrow}
\def\lsim{\:\raisebox{-0.5ex}{$\stackrel{\textstyle<}{\sim}$}\:}
\def\gsim{\:\raisebox{-0.5ex}{$\stackrel{\textstyle>}{\sim}$}\:}

 \begin{flushright}
{\large SHEP-03-07}\\
{\large TSL/ISV-2003-0271}\\
{\large August 2003}\\
\end{flushright}
\vspace*{5mm}
\begin{center}
{\Large\bf Pair production of charged Higgs bosons in association}\\
{\Large\bf  with bottom quark pairs at the Large Hadron Collider}\\[5mm]
{\Large S. Moretti\footnote{stefano@hep.phys.soton.ac.uk}}\\[1mm]
{\em School of Physics \& Astronomy, University of Southampton,\\
Highfield, Southampton SO17 1BJ, UK}
\\[5mm]
{\Large J. Rathsman\footnote{johan.rathsman@tsl.uu.se}}\\[1mm]
{\em High Energy Physics, Uppsala University,
Box 535, 751 21 Uppsala, Sweden}
\\[10mm]

\end{center}
\begin{abstract}
\noindent\small   
We study the process $gg\to b\bar b H^+H^-$ at large $\tan\beta$,
where it represents the dominant production mode of charged Higgs
boson pairs in a Type II 2-Higgs Doublet Model, including
the Minimal Supersymmetric Standard Model. The ability
to select this signal would in principle enable the measurements 
of some triple-Higgs couplings, which in turn would help
understanding the structure of the extended Higgs sector.
We outline a selection procedure that should aid in disentangling
the Higgs signal from the main irreducible background. This exploits
a signature made up by `four $b$-quark jets, two light-quark jets,
a $\tau$-lepton and missing energy'.
While, for $\tan\beta\gsim30$ and over a significant $M_{H^\pm}$ 
range above the top mass, a small signal emerges already at the 
Large Hadron Collider after 100 fb$^{-1}$, ten times
as much luminosity would be needed to perform accurate measurements
of Higgs parameters
in the above final state, rendering this channel a primary candidate
to benefit from the so-called
`Super' Large Hadron Collider option, for which a tenfold
increase in instantaneous luminosity is currently being considered.
  
\end{abstract}


\newpage

\section{Introduction}
\label{sec:intro}
Charged Higgs bosons
appear in the particle spectrum of a general 2-Higgs Doublet Model
(2HDM). We are concerned here with the case of a Type II 2HDM 
\cite{guide}, possibly in presence of minimal Supersymmetry (SUSY), the 
combination of the two yielding the so-called Minimal 
Supersymmetric Standard Model (MSSM). To stay with the Higgs sector
of the extended model, unless two or more neutral Higgs states\footnote{Of the
initial eight degrees of freedom pertaining to a complex Higgs
doublet, only five survive as real particles upon Electro-Weak  
Symmetry Breaking (EWSB), labeled as $h^0, H^0$, $A^0$
(the first two are CP-even or `scalars' whereas the third is CP-odd or
`pseudoscalar') and $H^\pm$, as three degrees of freedom 
are absorbed into the definition
of the longitudinal polarisation for the gauge bosons $Z^0$ and $W^\pm$,
upon their mass generation after EWSB.} 
are detected at the Large Hadron Collider (LHC), 
only the discovery of a spinless charged Higgs
state would unquestionably confirm the existence of new physics beyond
the Standard Model (SM),
since such a field has no SM counterpart. In the MSSM, e.g., 
if  $M_{H^\pm}, M_{A^0}, M_{H^0}\gg M_{h^0}$
and $\tan\beta$ is below 10 or so, the only
available Higgs state ($h^0$) is indistinguishable from the one of the SM:
this is the so-called `decoupling scenario'\footnote{One of
the Higgs masses, usually $M_{A^0}$ or $M_{H^\pm}$, and the 
ratio of the Vacuum Expectation Values (VEVs) of the up-type and 
down-type Higgs doublets (denoted by 
$\tan\beta$) are the two parameters that uniquely define
the MSSM Higgs sector at tree-level.}.

Not surprisingly then, a lot of effort has been put lately, by 
theorists and experimentalists alike, in
clarifying the Higgs discovery potential of the LHC in the charged
Higgs sector~\cite{reviewHpm}. (This is particularly true within the 
MSSM scenario, where one could also exploit interactions between
the Higgs and sparticle sectors~\cite{SUSY} in order to extend
the reach of charged Higgs bosons at the LHC, beyond the standard channels.)
Results are now rather encouraging, as charged Higgs bosons
could indeed provide the key to unveil the nature of EWSB over a
large area in $M_{H^\pm}$ and $\tan\beta$, as they
may well be the next available Higgs boson states, other than the $h^0$,
provided $\tan\beta$ is rather large (above 10 or so).
Once the $H^\pm$ and $h^0$ Higgs bosons will have been detected, the
next step would be to determine their interactions with SM particles,
among themselves and also with the other two neutral Higgs states,
$H^0$ and $A^0$. 
While the measurement of the former would have 
little to teach us as whether one is in presence of a general Type II 2HDM
or indeed the MSSM, constraints on the latter two would certainly help 
to clarify the situation in this respect. In fact, triple-Higgs vertices enter
directly the functional form of the extended Higgs potential and, once
folded within a suitable Higgs production process, may lead to the
measurement of fundamental terms of the extended model Lagrangian. As the
$H^\pm$ states have a finite Electro-Magnetic (EM) charge, the first
Lagrangian term of relevance would be the one involving two such states and a
neutral Higgs boson: chiefly, the vertices $H^+H^-\Phi^0$, where
$\Phi^0=h^0$ and $H^0$.\footnote{We are here only considering 
CP-conserving extensions of the SM Higgs sector such that there is
no ``$H^+H^-A^0$-vertex''.}
This requires the investigation of hard scattering 
processes with two charged Higgs bosons in the final state, as their
direct couplings to valence quarks in the proton would be very small, 
hence inhibiting processes like: e.g., 
$q\bar q'\to H^{\pm *}\to H^\pm\Phi^0$.

\section{Hadroproduction of charged Higgs boson pairs}
\label{sec:HH}

\begin{figure}
\begin{center}
\hskip-5.0cm\epsfig{file=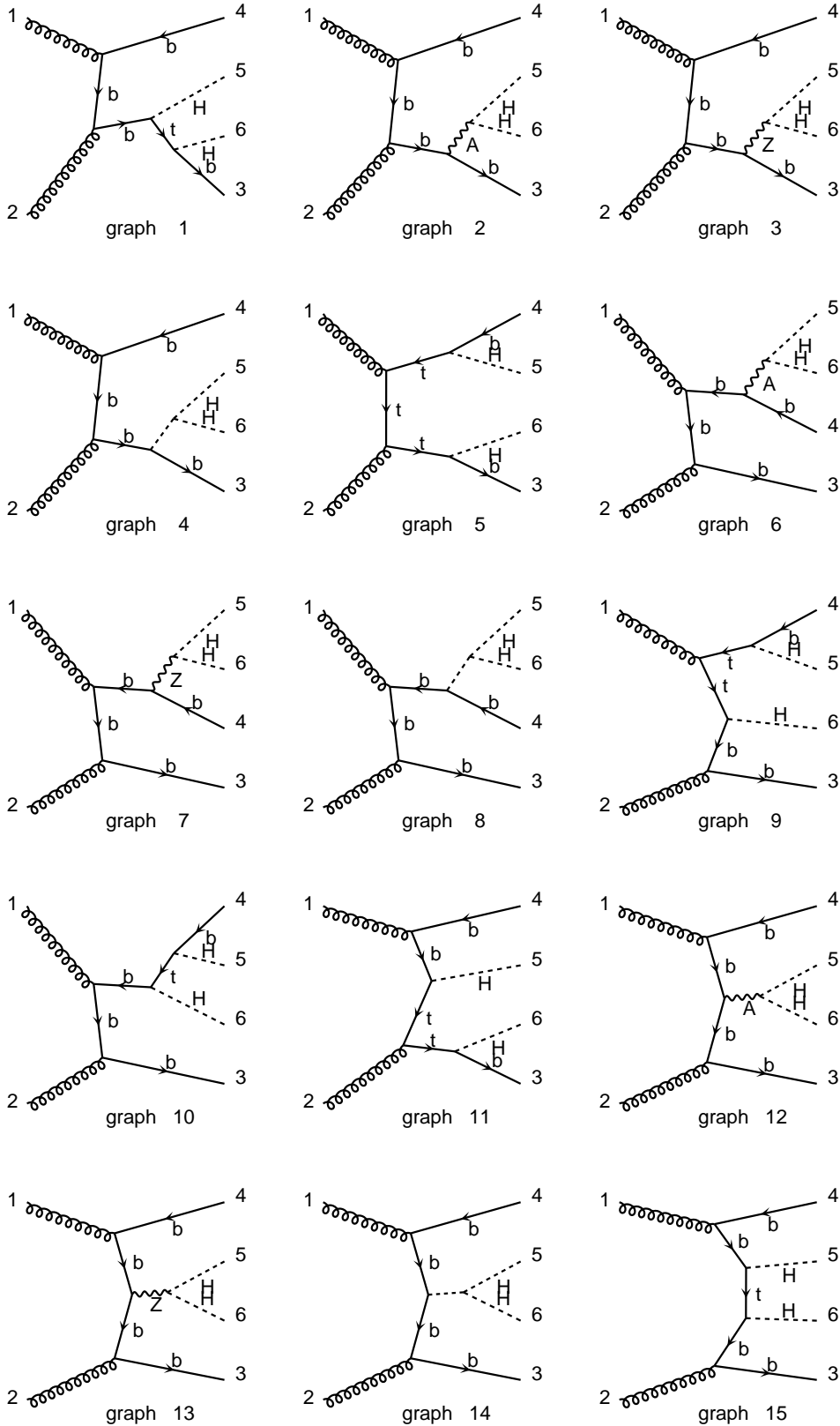,width=14cm,angle=0}
\end{center}
\vspace{-2.25cm}
\caption{Feynman diagrams in the unitary gauge for process (\ref{signal}).}
\label{fig:diagrams}
\end{figure}

A summary of all possible production modes of charged Higgs boson
pairs at the LHC in the MSSM 
can be found in Ref.~\cite{qqqqHH}.  Three channels dominate
$H^+H^-$ phenomenology at the LHC: (i) 
$q\bar q\to H^+H^-$
(via intermediate $\gamma^*/Z^{0*}$ production
but also via Higgs-strahlung off incoming $b\bar b$ pairs)~\cite{qqHH}; (ii) 
$gg\to H^+H^-$ (primarily
via a loop of top and bottom (s)quarks)~\cite{ggHH}; 
(iii) $ qq\to qq H^+H^-$ (via vector boson fusion)~\cite{qqqqHH}. 
Corresponding cross sections are found in Fig.~2 of Ref.~\cite{qqqqHH}.
For all phenomenologically relevant $\tan\beta$ values it is essentially 
the first process which dominates. One important aspect should be noted
here though, concerning the simulation
of the $b\bar b$ component of the $q\bar q\to H^+H^-$ process,
which can become the dominant contribution to the cross section
of process (i) at very large $\tan\beta$ values. In fact,
the use of a `phenomenological' $b$-quark parton density, as
available in most Parton Distribution Function
(PDF) sets currently on the market, requires crude 
approximations of the partonic kinematics, which result in a
mis-estimation of the corresponding contribution to the 
total production cross section. (The problem is well
known already from the study of the leading production processes
of charged Higgs bosons at the LHC, namely,
$\bar b g \rightarrow \bar tH^+$ and $gg\to b\bar t H^+$:
see, e.g.,~\cite{Borz,3b}.) In practice, the $b$-(anti)quark in
 the initial state 
comes from a gluon in the proton beam splitting into a collinear
$b\bar b$-pair, resulting in large factors of $\sim\alpha_{\rm S} 
\log(\mu_F/m_b)$,
where $\mu_F$ is the factorisation scale. 
These terms are 
then re-summed to all orders, $\sum_{n}\alpha_{\rm S}^{n} \log
^{n}(\mu_F/m_b)$, in evaluating the phenomenological $b$-quark PDF.
In contrast, in using a gluon density while computing the `twin'
process (iv) $gg\to b\bar b H^+H^-$
 (see Fig.~\ref{fig:diagrams}
for the associated Feynman graphs), one basically only includes the first 
terms ($n=1$) of
the corresponding two series, when the $b$ and $\bar b$ in the final state are
produced collinearly to the incoming gluon directions. It turns out that,
for $\mu_F\gg m_b$, as it is the case here if one uses the standard 
choice of factorisation scale $\mu_F\gsim 2M_{H^\pm}$,
the re-summed terms are large and over-compensate the contribution of 
the large transverse momentum
 region available in the gluon-induced case. In the end,
differences between the two cross sections as large as one
 order of magnitude are found, well
in line with the findings of Refs.~\cite{Borz} and~\cite{3b}, if
one considers that two $g\to b\bar b$ splittings are involved here.

One way to reconcile the large differences in the cross section
for the two processes,  $gg\to b\bar b H^+H^-$ and $ b\bar b \to H^+H^-$,
is to use a significantly lower factorisation scale, as argued 
in~\cite{Plehn:2002vy,Maltoni:2003pn,Boos:2003yi,Harlander:2003ai} 
for similar processes. 
Following the suggestion in part A.1.~of~\cite{Cavalli:2002vs}, we  
look at the transverse momentum
distribution of the $b$-quarks in the process $gg\to b\bar b H^+H^-$,
as shown in Fig.~\ref{fig:fact}, to
get an indication of the most suitable factorisation scale for
$ b\bar b \to H^+H^-$.  From the figure we see that a proper choice
for the latter, when $M_{H^\pm}=215$ GeV, is of the order
$\mu_F=0.1\sqrt{\hat{s}}\simeq40$ GeV (at this point the distribution 
reaches about half of its ``plateau'' value\footnote{This number
is not too dissimilar from the one recommended in~\cite{Boos:2003yi}
on the basis of the same argument applied to
the single $H^\pm$ production mode $gg\to b\bar t H^+$,
as $M/4$, where $M$ is the `threshold mass' $m_t+M_{H^\pm}$.}) 
rather than, e.g., $\mu_F=\sqrt{\hat{s}}$.
Using such a lower scale we do get a much better agreement
between the leading order (LO) cross sections for the two processes, 
as shown in Tab.~\ref{tab:fact} 
in the case of the MSSM specified below ($M_{A^0}=200$ GeV and 
$\tan\beta=30$) if the renormalisation scale ($\mu_R$) is also changed 
accordingly. However, one should also bear in mind that both processes
are subject to possibly large QCD corrections and that the choice of
(factorisation and/or renormalisation)
scales that minimises the differences between 
the two descriptions in higher orders of 
$H^+H^-$ production may alternatively
be viewed as the most suitable one. Or else, one may arguably choose a
scale that minimises the size of the higher order corrections themselves
in either process independently of the other. All such additional values 
may eventually turn out to be different from the one extracted from
Tab.~\ref{tab:fact}. Such exercises in higher orders  cannot 
unfortunately be performed in the present context, as next-to-leading order 
(NLO) corrections to the two processes of interest are unavailable. Yet, some
guidance may be afforded again by the study of the single charged
Higgs production modes already referred to. 
In fact, NLO corrections to $\bar b g\to \bar t H^+$ were first 
computed in Ref.~\cite{Zhu} and then later confirmed in~\cite{Plehn:2002vy}.
Following Ref.~\cite{Plehn:2002vy}, it is clear that a choice for the
renormalisation scale $\mu_R$ as low as the one recommended for
the factorisation one $\mu_F$ is not sustainable
for $\bar bg\to tH^+$ at NLO, no matter the choice
for the latter: see Fig.~5 of
~\cite{Plehn:2002vy}. Besides, if one fixes, e.g., $\mu_R=(m_t+M_{H\pm})/2$
but varies $\mu_F$, the minimal difference between the NLO and the LO
results for  $\bar bg\to tH^+$ is found at  large $\mu_F$, at
values around or even larger than $m_t+M_{H\pm}$ (again, see
Fig.~5 of Ref.~\cite{Plehn:2002vy}). Be the most suitable combination of 
scales as it may, we take here a pragmatical attitude and use the standard 
setup $\mu_F=\mu_R=\sqrt{\hat{s}}$ throughout, as this corresponds to 
the most conservative choice in terms of the overall normalisation 
for $gg\to b\bar b H^+H^-$ (Tab.~\ref{tab:fact}) -- as it becomes minimal -- and
keeping in mind that its cross section can be up to a factor $\sim5$
larger depending on the choice of factorisation and renormalisation scales. 

\begin{table}[t]
\caption{Cross sections for $gg\to b\bar b H^+H^-$ and 
$ b\bar b \to H^+H^-$ as functions of the factorisation ($\mu_F$) and 
renormalisation ($\mu_R$) scales (at leading order the 
$ b\bar b \to H^+H^-$ cross section does not depend on $\mu_R$) in the
MSSM specified below ($M_{A^0}=200$ GeV and $\tan\beta=30$). }
\begin{center}
\begin{tabular}{cccc}
\hline
$\mu_F$         & $\mu_R$         & $\sigma(gg\to b\bar b H^+H^-)$ [fb]&  $\sigma(b\bar b \to H^+H^-)$   [fb]  \\
$\sqrt{\hat{s}}$ & $\sqrt{\hat{s}}$ &  1.4  &   \\
$\bra m^T_b \ket^\dagger$ & $\sqrt{\hat{s}}$ &  2.3  &   \\
$\bra m^T_b \ket^\dagger$ & $\bra m^T_b \ket^\dagger$ &  8.2  &   \\
$\sqrt{\hat{s}}$ &   &   &  7.5 \\
$0.1\sqrt{\hat{s}}$ &            &      &  4.4 \\
\hline
\end{tabular}
\end{center}
{$^\dagger$ Here, the exact definition is:
$\bra m^T_b \ket=\sqrt{m_b^2+((p^T_{b})^2+(p^T_{\bar{b}})^2)/2}$.}
\label{tab:fact}
\end{table}

Under any circumstances, a clear message that emerged from NLO computations 
of $\bar bg\to tH^+$ with respect to
the LO ones of $gg\to b\bar t H^+$ is that
the former (duly incorporating a running NLO $b$-quark mass in the Yukawa
coupling to the charged Higgs boson) agree better with the latter if these
use the pole $b$-quark mass instead, see Fig.~4 of
Ref.~\cite{Plehn:2002vy}. By analogy, in the reminder of our
paper, we will make the same assumption (of a pole $b$-quark mass
entering the $b\bar t H^+$ vertex) in our $gg\to b\bar b H^+H^-$ 
process at LO. Finally, 
while a well defined procedure exists in order to compute both inclusive
and exclusive cross sections when
combining the $b\bar b$- and $gg$-initiated processes, through the subtraction
of the common logarithm terms~\cite{Borz} and/or by a cut in phase
space~\cite{Jaume}, it should be noticed 
that process (iv) is the only contributor when one exploits the tagging
of both the two $b$-quark jets produced in association with the charged Higgs
boson pair.

It is precisely the intention of this note that of pursuing a similar
strategy in order to extract a possible $b\bar b H^+H^-$ signal,
as it has already been shown the beneficial effects of triggering on the
`spectator' $b$-jet in the $gg\to t\bar b H^-$ case, in order to
improve on the discovery reach of charged Higgs bosons at the LHC~\cite{4b}.
Furthermore, if vertices of the type $H^+H^-\Phi^0$ are to be studied 
experimentally, one should appreciate the importance of the $gg\to b\bar b
\Phi^0$ subprocess (from which two charged Higgs bosons would stem out
of the above triple-Higgs coupling: 
see diagrams $4, 8, 14, 19, 23, 29, 34$ and 38
in Fig.~\ref{fig:diagrams}) 
by recalling that the latter reaction is the dominant
production mode of neutral Higgs scalars (chiefly, the $H^0$
state) at $\tan\beta$ values above 7 or so, for any neutral
Higgs mass of phenomenological interest~\cite{Spira}, this justifying our
choice of privileging here the $gg\to b\bar b H^+H^-$ channel. Finally,
whereas the already recalled MSSM relation 
$M_{H^\pm}\simeq M_{A^0}\simeq M_{H^0}\gg M_{h^0}$ 
clearly prevents the appearance of a 
$H^0\to H^+H^-$ resonance
in the diagrams proceeding via intermediate states of the form 
$gg\to b\bar b H^0$, this is no longer true in a general 2HDM,
wherein one may well have $M_{H^0}>2 M_{H^\pm}$, with the consequent
relative enhancement of the mentioned subset of diagrams with respect 
to all others appearing in Fig.~\ref{fig:diagrams}.

We will attempt the signal selection for the case of rather heavy charged
Higgs bosons, with masses above that of the top quark. The case for 
the existence of such massive Higgs 
states has in fact become phenomenologically pressing,
since rumours of a possible evidence of light charged Higgs bosons being
produced at LEP2~\cite{HpmExcessLEP} have faded away. 
Instead, one is now left from LEP2 
with a model independent limit on $M_{H^\pm}$, of order $M_{W^\pm}$.
Within the MSSM, 
the current lower bound on a light Higgs boson
state, of approximately 110 GeV~\cite{HLEP2}, 
can be converted at two-loops into a minimal
value for the charged Higgs boson mass, of order 130--140 GeV, for
$\tan\!\beta \, \simeq \,3$--$4$\footnote{Recall that the tree-level
relation between the masses of the charged and pseudoscalar 
Higgs boson, $M_{H^{\pm}}^2 = M_{A^0}^2 
+ M_{W^\pm}^2$, is almost invariably quite 
insensitive to higher order corrections~\cite{mhc-cor}.}. This bound
grows rapidly stronger as $\tan\!\beta$ is decreased while tapering
very gradually as $\tan\!\beta$ is increased  (staying in the
$110$--$125\, \hbox{GeV}$ interval for $\tan\beta \gsim7$). Besides, 
in the mass interval $M_{H^\pm}< m_t$, charged Higgs bosons 
could well be found at Tevatron (Run 2)~\cite{Run2}, which has 
already begun data taking at $\sqrt s_{p\bar p}=
2$ TeV at FNAL, by exploiting their production in top decays,
$t\to b H^+$, and the tau-neutrino detection mode,
 $H^-\to \tau\bar\nu_\tau$. In contrast, if $M_{H^\pm}\gsim~m_t$ 
(our definition of a `heavy' charged Higgs boson), one will necessarily
have to wait for the advent of the LHC, $\sqrt s_{pp}=14$ TeV, at CERN.
As hinted
at in the beginning, we also make the assumption in our study 
that the charged Higgs
boson mass is already known, e.g., from studies of the leading production
and decay channels, $gg\to t\bar b H^-$ and $H^-\to\tau^-\bar\nu_\tau$
or $b\bar t$, during the first years of running of the CERN hadron machine.

Under the above parameter assumptions, i.e., large $\tan\beta$ 
($\gsim10$) and large
$M_{H^\pm}$ values ($\gsim m_t$), a sensible choice of decay
channels~\cite{BRs} 
for our pair of charged Higgs bosons would be to require one
to decay via the leading mode, i.e., $H^+\to t\bar b$ (with
the $t$-quark further decaying hadronically, so to allow for the
kinematic reconstruction of the charged Higgs boson resonance
in a four-jet system) and the other via
$H^-\to\tau^-\bar\nu_\tau$ (whose rate increases with $\tan\beta$
and that yields a somewhat cleaner trigger in the LHC environment,
independently of whether the $\tau$-lepton decays leptonically
or hadronically, as opposed to the above multi-jet and high hadron-multiplicity
signature). 
As such decays would induce an intermediate signal state made
up by $b\bar b t\bar b \tau^-\bar\nu_\tau$ and since
we will assume tagging all four $b$'s, it is clear that
the dominant irreducible background would be 
$b\bar b t\bar t$ production followed by the decay $\bar t\to
\bar b W^-\to\bar b \tau^-\bar\nu_\tau$.

\section{Calculation}
\label{sec:calcul}

The hard subprocess describing our signal is then
\begin{equation}\label{signal}
gg\to b\bar b H^+ H^-,
\end{equation}
whereas for the main irreducible background we have to deal with 
\begin{equation}\label{background}
gg\to b\bar b t\bar t.
\end{equation}
(We neglect here the computation of the quark-antiquark initiated
components of both signal and background, i.e.,
$q\bar q\to b\bar b H^+ H^-$ and
$q\bar q\to b\bar b t\bar t$, respectively,
as they are negligible at the LHC, in comparison to the
gluon-gluon induced modes.) The 
matrix elements for (\ref{signal})--(\ref{background})
have been calculated by using the HELAS~\cite{HELAS}
subroutines and MadGraph~\cite{Tim}. 
All unstable particles
entering the two processes ($t, H^\pm$ and $W^\pm$) were generated
not only off-shell (i.e., with their natural widths) but also in 
Narrow Width Approximation (NWA) for comparison. 
For the MSSM and 2HDM Higgs
bosons, the program {\tt HDECAY}~\cite{HDECAY} has been exploited to
generate the decays rates eventually used in the Monte Carlo (MC)
simulations. For the MSSM, we have assumed the following set up
for the relevant SUSY input parameters: $\mu=0$, $A_\ell=A_u=A_d=0$ (with
$\ell=e,\mu,\tau$ and $u/d$ referring to $u/d$-type quarks) and 
$M_{\rm{SUSY}}=1$ TeV, the latter
implying a sufficiently heavy scale for all sparticle masses, so
that no $H^\pm\to$ SUSY decay can take place\footnote{The only 
possible exception in this mass hierarchy would be the
Lightest Supersymmetric Particle (LSP), whose mass may well be smaller
than the $M_{H^\pm}$ values that we will be considering, a state 
in which however a charged Higgs boson cannot decay to, as $R$-parity and
EM charge conservation would require the presence of an additional heavy
chargino.}.

As a 2HDM configuration, we have basically maintained the previous
setup in the relevant input parameters of {\tt HDECAY}, with
the most important difference being the 
 assumption of a different relation between the
input $M_{A^0}$ value and the derived $M_{H^0}$ one, 
by 
assuming a linear relation between the $H^0$ and $H^\pm$ masses,
i.e., $M_{H^0}=x~  H^\pm$, with $x$ being a number larger than 2,
while maintaining the MSSM relations among the $H^\pm$ and $A^0$ masses,
thereby allowing for
the already intimated onset of $H^0\to H^+H^-$ 
resonant decays in diagrams $4, 8, 14, 19, 23, 29, 34$ and 38 of
Fig.~\ref{fig:diagrams}. As already remarked upon, this is
a crucial phenomenological
difference with respect to the MSSM, wherein such a decay threshold is
never reached over the unexcluded region of parameter space. 
Another important difference, in a more general 2HDM, is the value of 
the triple Higgs coupling $g_{H^0H^+H^-}$ which can be much larger 
than what is the case in the MSSM. 

Before giving the details of the 2HDM setup we are using let us recall the
most general CP-conserving 2HDM scalar potential which is symmetric under 
$\Phi_1\to-\Phi_1$  up to softly breaking
dimension-2 terms (thereby allowing for loop-induced flavour 
changing neutral currents)~\cite{guide},
\begin{eqnarray}\label{eq:potential}
V(\Phi_1,\Phi_2) &=& 
\lambda_1(\Phi_1^\dagger\Phi_1-v_1^2)^2
+\lambda_2(\Phi_2^\dagger\Phi_2-v_2^2)^2
 \nonumber \\ &&
+\lambda_3\left[(\Phi_1^\dagger\Phi_1-v_1^2)+(\Phi_2^\dagger\Phi_2-v_2^2)\right]^2
 \nonumber \\ &&
+\lambda_4\left[(\Phi_1^\dagger\Phi_1)(\Phi_2^\dagger\Phi_2)-
                (\Phi_1^\dagger\Phi_2)(\Phi_2^\dagger\Phi_1)\right]
 \nonumber \\ &&
+\lambda_5\left[{\mathrm{Re}}(\Phi_1^\dagger\Phi_2)-v_1v_2\right]^2
+\lambda_6\left[{\mathrm{Im}}(\Phi_1^\dagger\Phi_2)\right]^2
\end{eqnarray}
where $v_1^2+v_2^2=v^2=2M_W^2/g^2\simeq (174$ GeV)$^2$.
In general, the potential is thus parameterised by seven parameters
(the $\lambda_i$ and $\tan\beta=v_2/v_1$) whereas in the MSSM only
two of them are independent.  In the following we will
replace five of the $\lambda_i$ with the masses of the Higgs 
bosons ($M_{h^0},M_{H^0},M_{A^0},M_{H^\pm}$)
and the mixing angle $\alpha$ of the CP-even Higgs bosons.

From the scalar potential the different three- and four-Higgs couplings
can be obtained. (See~\cite{Boudjema:2001ii,Gunion:2002zf} for a 
complete compilation of couplings in a general CP-conserving 2HDM.)
Using the Higgs masses and $\alpha$ as parameters together with $\lambda_3$
the $g_{H^0H^+H^-}$ coupling takes a particularly simple form
(see for example~\cite{Arhrib:1998gr})
\begin{eqnarray}\label{eq:2hdmcoupling}
g_{H^0H^+H^-} & = & 
-i \frac{g}{M_W}
\left[
\cos(\beta-\alpha)\left(M_{H^\pm}^2-\frac{M_{H^0}^2}{2}\right)
+\frac{\sin(\alpha+\beta)}{\sin2\beta}\left\{\frac{1}{2}(M_{H^0}^2+M_{h^0}^2)+
\right.\right. \nonumber \\ && \left. \left.
 + 4\lambda_3v^2
  -\frac{1}{2\sin2\beta}\left[ 
    \sin2\alpha+2\sin(\alpha-\beta)\cos(\alpha+\beta)\right](M_{H^0}^2-M_{h^0}^2)
    \right\}
\right] \: .
\end{eqnarray}
For the other three- and four-Higgs couplings we refer 
to~\cite{Boudjema:2001ii,Gunion:2002zf}. 

In the 2HDM  that we will adopt below we will
start from the MSSM parameter values for 
$M_{H^0},M_{A^0},M_{h^0},M_{H^\pm},\tan\beta,\alpha,(\lambda_5-\lambda_6)/2 $ 
using $M_{A^0}$ and $\tan\beta$ as input.
As already mentioned we then change from $M_{H^0}^{\rm MSSM}$ to 
$M_{H^0}^{\rm 2HDM}=2.2 M_{H^\pm}$ and in addition 
$M_{h^0}=1.7 M_{H^\pm}$ while
keeping the other Higgs boson masses and $\tan\beta$ fixed. The
choice $M_{h^0}=1.7 M_{H^\pm}$ has been found to be favourable in 
order to get a larger $g_{H^0H^+H^-}$ coupling and at the same time
avoid negative interference between $H^0$ and $h^0$ resonances.
The remaining two 
parameters, $\alpha$ and $(\lambda_5-\lambda_6)/2$, are then determined by
randomly picking one million ($\alpha,(\lambda_5-\lambda_6)/2$)-points,
in the ranges $[-\pi/2,\pi/2]$ and $[-4\pi,4\pi]$ respectively,
and keeping the one which gives the largest effective coupling,
$g_{H^0H^+H^-}\cos\alpha$, thereby also taking into account the 
$H^0b\bar b$-coupling. In order to accept a point we also check 
that the following conditions are fulfilled: the potential is bounded
from below, the $\lambda_i$ fulfill the unitarity 
constraints~\cite{Akeroyd:2000wc}, the contribution 
to $\Delta\rho < 10^{-3}$ (although with the above setup for the 
Higgs masses we are more or less guaranteed not to violate any 
experimental bounds on the $\rho$-parameter~\cite{guide}), 
and the combined partial width for the
three $H^0 \to h^0h^0, A^0A^0, H^+H^-$ decays is smaller than $M_{H^0}/2$.
(We have checked that the partial widths 
of the $H^0 \to h^0h^0h^0, h^0A^0A^0, h^0H^+H^-$ decay channels are negligible.)

Some examples of the actual values of the $\lambda_i$ we use in this 
study are given in Tab.~\ref{tab:lambda}\footnote{The selection 
procedure outlined above did not lead to any acceptable solution 
for $M_{A^0} \gsim 310$ GeV.} together with the corresponding values of
$\alpha$ and $g_{H^0H^+H^-}$. 
From the table one sees that the effective coupling $g_{H^0H^+H^-}\cos\alpha$
decreases quite rapidly as $M_{A^0}$ increases, giving a correspondingly
smaller cross-section.(This will be shown in more detail below.) However,
it should be kept in mind that the 2HDM setup we are using is not the
most general one based on the scalar potential (\ref{eq:potential}) and
that there may be other parts of the parameter space which 
we have not scanned that give a larger cross section. At the same time it 
should be said that we have already tried different relations between the 
Higgs masses other than the ones given above, yet we have not made any 
further detailed investigations.

\begin{table}[t]
\caption{ Examples of values of the parameters in the Higgs potential used 
for different values of $M_{A^0}$ together with the corresponding values of
$\alpha$ and $g_{H^0H^+H^-}$.}
\begin{center}
\begin{tabular}{c|cccccc|cc}
\hline
\hline
$M_{A^0}$   & $\lambda_1$   & $\lambda_2$   & $\lambda_3$   & $\lambda_4$   & $\lambda_5$   & $\lambda_6$ & $\alpha$   & $g_{H^0H^+H^-}$ \\
$[$GeV$]$   &               &               &               &               &               &             &            & $[$GeV$]$ \\
\hline
150 & $-$4.81618 & $-$4.02795 &   4.81618  &  0.941863 &   4.13114 &  0.745433  & 0.5064  &  $-$1235 \\
200 & $-$3.90929 & $-$2.75838 &   3.90929  &   1.52164 &   7.16375 &   1.32521  & 0.2686  &  $-$563 \\
250 &   1.69717  &   3.17924  & $-$1.54105 &   2.26707 &   10.9783 &   2.07065  & 0.0363  &      76 \\
300 &   5.16558  &   7.63614  & $-$3.79044 &   3.17816 &   9.19537 &   2.98173  & -1.539  &  $-$1078 \\
\hline
\hline
\end{tabular}
\end{center}
\label{tab:lambda}
\end{table}

To get an explicit example of the large differences between the more 
general 2HDM we are considering 
and the MSSM, we compare the two using $M_{A^0}=200$ GeV and $\tan\beta=30$ 
as input values, as we will do below.  In this case 
we get $g_{H^0H^+H^-}^{\rm 2HDM}=-563$ GeV in the 2HDM 
instead of the MSSM value,
$g_{H^0H^+H^-}^{\rm MSSM}=-1.8$ GeV. 
With the value of $\alpha^{\rm 2HDM}=0.26859$ not being much larger than 
$\alpha^{\rm MSSM}=-0.05774$ the difference in effective coupling 
$g_{H^0H^+H^-}\cos\alpha$ is more than a factor hundred and as we will
see below it has a large impact on the magnitude of the cross section.
(The $H^0$ widths will of course be different too in the
MSSM and 2HDM just described: their effects have been included in the
numerical analysis.)

As intimated already in Sect.~\ref{sec:HH}, 
a non-running $b$-quark mass was adopted for both the kinematics and the 
Yukawa couplings: $m_b=4.25$ GeV. For the top parameters, we have
taken $m_t=175$ GeV with $\Gamma_t$ computed according
to the model used (in the limit $M_{H^\pm}\gg m_t$,
we have $\Gamma_t=1.56$ GeV in both the 2HDM and MSSM scenarios
considered here). EW parameters were as follows: 
$\alpha_{\rm{EM}}=1/128$, $\sin^2\theta_W=0.232$, with
$M_{Z^0}=91.19$ GeV ($\Gamma_{Z^0}=2.50$ GeV)
and 
$M_{W^\pm}=M_{Z^0}\cos\theta_W$ ($\Gamma_{W^\pm}=2.08$ GeV).
For the $\tau$-lepton mass we used $m_\tau=1.78$ GeV, whereas all
the other leptons and quarks were assumed to be massless.

The integrations over the multi-body final states 
have been performed numerically with the aid of VEGAS~\cite{VEGAS},
Metropolis~\cite{Metro} and {\tt 
RAMBO}~\cite{RAMBO}, for checking purposes. Finite calorimeter resolution
has been emulated through a Gaussian smearing in transverse momentum,
$p^T$, with $(\sigma(p^T)/p^T)^2=(0.60/\sqrt{p^T})^2 +(0.04)^2$ for all 
jets  and   $(\sigma(p^T)/p^T)^2=(0.12/\sqrt{p^T})^2 +(0.01)^2$ for 
leptons. The corresponding missing transverse momentum, 
$p^T_{\mathrm{miss}}$, was reconstructed from the vector sum of 
the visible momenta after resolution smearing. Furthermore, we 
have identified jets with the 
partons from which they originate and applied all cuts directly to the
latter, since parton shower and hadronisation were not included in our
study. The only exception is the $\tau$-lepton decay which has been
taken into account using the {\sc Pythia}~\cite{Sjostrand:2000wi}
 MC event generator.

 As default PDFs 
we have adopted the set MRS98LO(05A)~\cite{DURPDG}
with $Q=\mu=\sqrt{\hat{s}}$ as factorisation/renormalisation
scale for both signal and background. The same scale entered
the evolution of $\alpha_{{\rm S}}$, which was performed at one-loop,
with a choice of $\Lambda_{\rm{QCD}}^{n_f=4}$ consistent with the PDF set
adopted. In fact, we have verified that the spread in the inclusive
cross sections, for both signal and background, as obtained by using the
five different parameterisations of MRS98LO and also CTEQ4L~\cite{DURPDG}
was within $6-7\%$ of the values quoted for MRS98LO(05A) in the remainder
of the paper.

\section{Selection}
\label{sec:sele}

The signature that we are then considering is in practice:
\begin{equation}\label{signature}
4~b{\rm{-jets}}~+~2~{\rm{light-jets}}~+~\tau~+~p^T_{\rm{miss}},
\end{equation}
wherein the two $b$-jets from the hard process 
are actually defined as the two most
forward/backward ones that also display a displaced vertex.

We have assumed a standard detector configuration by imposing 
acceptance and separation cuts on {\sl all} light-quark (including
$c$'s) and $b$-jets, labeled as $j$ and $b$, respectively, as follows:
\begin{equation}\label{cuts1}
 p^T_{b} >  20    ~{\rm{GeV}},\qquad
 |\eta_{b}| <  2.5 ,\qquad
 p^T_{j} >  20    ~{\rm{GeV}},\qquad
 |\eta_{j}| <  5,\qquad
 \Delta R_{jj,jb} >  0.4. 
\end{equation}
The two most forward/backward $b$-jets (with pseudorapidities
of opposite sign) are further required to yield 
\begin{equation}\label{cuts2}
 M_{bb}    >  M_{H^\pm}
\end{equation}
for their invariant mass. 
For $\tau$-jets (we only consider hadronic decay modes) we impose:
\begin{equation}\label{cuts3}
 p^T_{\tau} >  10    ~{\rm{GeV}},\qquad\qquad
 |\eta_{\tau}| <  2.5,\qquad\qquad
 \Delta R_{j\tau,b\tau} >  0.4.
\end{equation}
The setup corresponds to the standard ATLAS/CMS 
detectors. Presently it is not clear to what extent this setup
will also be applicable for the same apparata in the context of the 
Super LHC (SLHC) option~\cite{SLHC}.

Having now excluded the two most forward/backward $b$-jets from
the list of jets, we impose hadronic $W^\pm$- and $t$-mass reconstruction:
\begin{equation}\label{cuts4}
 |M_{jj}-M_{W^\pm}|    <  15    ~{\rm{GeV}},\qquad\qquad
 |M_{bjj}-m_t|   <  35    ~{\rm{GeV}},
\end{equation}
where the two light-quark jets entering the last inequality are
of course the same fulfilling the first one. 
Finally, the missing transverse momentum should 
be\footnote{For simplicity we have kept this cut fixed, whereas in 
a more detailed analysis one would preferentially make the cut depend
on the mass of the charged Higgs boson.}:
\begin{equation}\label{cuts5}
 p^T_{\rm{miss}} >  60 ~{\rm{GeV}}.
\end{equation}

The combined effects of these cuts on the signal (\ref{signal}) in the 
MSSM and 2HDM models is shown in Figs.~\ref{fig:bbHH_cuts} and 
\ref{fig:newbbHH_cuts} respectively in the case of retaining the finite
width effects using off-shell masses for the $H^\pm$ and $W^\pm$. 
(The difference when instead using the NWA is very small and this 
option is therefore not shown.) As can be seen from the figures the effects
of the cuts on the magnitude of the cross sections is quite severe. On the
other hand the cuts are needed (especially in case of the MSSM) in order
to beat down the background as is illustrated in Fig.~\ref{fig:compare_cuts}.
From the figure it is also clear that the signal cross section
reaches its maximum around $M_{H^\pm}=200$ GeV and that the magnitude of the 
signal varies by up to 2 orders of magnitude depending on which model we
are considering. It should also be noted that
in the 2HDM setup we are using it was not possible to go above
$M_{H^\pm}\simeq 320$ GeV (corresponding to $M_{H^0}=700$ GeV)
due to the unitarity contraints~\cite{Akeroyd:2000wc}. 
In the following we will be considering the case $M_{H^\pm}\simeq 215$ GeV 
(corresponding to $M_{A^0}=200$ GeV) in more detail.

After the above cuts have been implemented and the jet
momenta assigned, one can reconstruct the would-be charged
Higgs boson mass, by pairing the three jets entering the equation
in the right-hand side of (\ref{cuts4}) with the left-over
central jet, a quantity which we denote by  $M_{4{\rm{jets}}}$. 
The corresponding mass spectrum is presented
in Fig.~\ref{fig:Hmass_cuts} 
for our two customary setups of MSSM and 2HDM, assuming $M_{H^\pm}=
215$ GeV and $\tan\beta=30$ as representative values.
The figure shows clear peaks at the charged Higgs boson mass for the signal
on top of a combinatorial background in both models
whereas for the background process there is no such peak. 
Thus, by selecting events with
\begin{equation}\label{M4cut}
180~\mathrm{GeV} < M_{4{\rm{jets}}} < 250~\mathrm{GeV}
\end{equation} 
we can get an additional discrimination against the background.

Furthermore, using the visible $\tau$-jet momentum 
and the missing transverse one,
it is possible to reconstruct the transverse mass, as 
\begin{equation}\label{MT}
M^T_{\tau\nu_\tau}=\sqrt{2 p_\tau p^T_{\rm{miss}} (1-\cos\Delta\phi)},
\end{equation}
with $\Delta\phi$ the relative angle between the two momenta, 
a quantity which is 
ultimately correlated to the actual value of the mass resonance
yielding $\tau^-\bar\nu_\tau$ pairs ($H^-$ in the signal and
$W^-$ in the background). We show this observable in 
Fig.~\ref{fig:Tmass_cuts} (where it is denoted by $M_T$), again,  
for our two customary setups of MSSM and 2HDM.
From the figure it is clear that the transverse mass for the signal
in the two models extends all the way out to $\sim M_{H^\pm}$.
Comparing with the background, which starts to drop quite drastically around 
$M^T_{\tau\nu_\tau}\approx M_{W^\pm}$, 
we see that it is advantageous to introduce a cut on the transverse 
mass of the order of 
\begin{equation}\label{MTcut}
M^T_{\tau\nu_\tau}> M_{H^\pm}/2~(\simeq 107 ~\mathrm{GeV}).
\end{equation}
This gives a very strong suppression of the background whereas the signal
is only mildly affected.

After having reconstructed the two charged Higgs bosons and applied the
respective cuts (\ref{M4cut}) and (\ref{MTcut}) we can form the 
(effectively, transverse) mass $M^T_{{\rm{4jets}}+\tau\nu }$ 
of the combined two charged Higgs boson system, giving the possibility 
to look for possible resonances. As can be seen from the magnitudes of the 
cross sections in Fig.~\ref{fig:HHmass_cuts} the selection outlined
above gives a very clear signal in the 2HDM case with essentially no 
background, whereas in the MSSM case the signal is still clear but very small. 
Of course, the cuts could be tightend to give a very clear signal also in 
the MSSM case, but the absolute cross section would then be even smaller. 
From the figure it is also clear that 
there is a significant difference in shape between the two models,
with the 2HDM showing a clear enhancement for  
$M^T_{{\rm{4jets}}+\tau\nu } \lsim 500$ GeV due to the resonant contributions.
However, given the limited statistics that will be available at the LHC,
it is not clear to what extent the difference in shape alone can be used to
extract any information about possible $H^0\to H^+H^-$ 
resonances and the corresponding coupling.
In addition the difference in shape will be smaller if the width of the 
$H^0$ Higgs boson is larger and/or the detector resolution is worse than 
what we have assumed.

\section{Conclusion}
\label{sec:summa}

As we have shown, it is possible to outline a selection procedure that
enables one to extract a signal of heavy charged Higgs pair production 
in association with two $b$-quarks at $\tan\beta\gsim30$ in extensions of
the standard model with two Higgs doublets of Type II. In a general case
the mass relations in the Higgs sector may be favourable such that a 
sizeable signal would appear already at the LHC through the resonant
channel $gg \to b\bar b H^0\to b\bar b H^+H^- $. However, in the MSSM the
resonance is not accessible over the allowed parameter region
and the non-resonant contributions turn
out to be very small making it difficult to extract a signal even after
upgrading the luminosity at LHC by a factor ten (SLHC).
The large difference in cross section between the MSSM and a more general
2HDM shows that indeed the pair production of charged Higgs production
is sensitive to the $H^0 H^+H^-$ coupling even though it will probably be
difficult to reconstruct a resonant transverse mass peak mainly due to the 
limited statistics and possibly also due to the finite detector resolution.

In our study we have not included effects of the $b$-tagging efficiency.
On the one hand, requiring four $b$-tags will give a sizeable reduction 
(of the order of a factor ten) of the signal as well as the backgrounds.
On the other hand, the selection procedure outlined above was designed
to get a signal in the case of the MSSM leading to unnecessarily tight
cuts for the general 2HDM case. Furthermore, 
being a leading order calculation, the cross section we get for the
signal is also sensitive to the choice of factorisation and 
renormalisation scales. If we, instead of using the standard choice of the
invariant mass of the hard subsystem, use the mean transverse
mass of the two $b$-quarks in the $gg \to b\bar b H^+H^- $ process as 
scale, the
cross section increases by a factor 5. Such a scale choice also gives a better
agreement with the cross section for the `twin' process 
$b\bar b \to H^+H^- $. In order to get a better handle 
on the uncertainties due to scale choices a next-to-leading
order calculation will eventually be necessary.
Nonetheless, we believe that our results already call for the attention 
of ATLAS and CMS in
further exploring the scope of the (S)LHC in reconstructing the
form of the Higgs potential in extended models through signals 
of charged Higgs boson pairs. Besides,
in presence of parton shower, hadronisation and detector effects,
one may also realistically attempt exploiting $\tau$-polarisation
techniques in hadronic decays of the heavy lepton \cite{Roy} in order
to increase the signal-to-background rates, an effort that was beyond the
scope of our parton level analysis.

\section*{Acknowledgments} SM is grateful to The Royal Society
of London (UK) for partial financial support in the form of a 
Study Visit and thanks the High Energy Theory Group
of the Department of Radiation Science of Uppsala University for
kind hospitality.

\def\pr#1 #2 #3 { {\rm Phys. Rev.}            {#1}, #3 (#2)}
\def\prd#1 #2 #3{ {\rm Phys. Rev. D}          {#1}, #3 (#2)}
\def\prl#1 #2 #3{ {\rm Phys. Rev. Lett.}      {#1}, #3 (#2)}
\def\plb#1 #2 #3{ {\rm Phys. Lett. B}         {#1}, #3 (#2)}
\def\npb#1 #2 #3{ {\rm Nucl. Phys. B}         {#1}, #3 (#2)}
\def\prp#1 #2 #3{ {\rm Phys. Rep.}            {#1}, #3 (#2)}
\def\zpc#1 #2 #3{ {\rm Z. Phys. C}            {#1}, #3 (#2)}
\def\epjc#1 #2 #3{ {\rm Eur. Phys. J. C}      {#1}, #3 (#2)}
\def\mpl#1 #2 #3{ {\rm Mod. Phys. Lett. A}    {#1}, #3 (#2)}
\def\ijmp#1 #2 #3{{\rm Int. J. Mod. Phys. A}  {#1}, #3 (#2)}
\def\ptp#1 #2 #3{ {\rm Prog. Theor. Phys.}    {#1}, #3 (#2)}
\def\jhep#1 #2 #3{ {\rm J. High Energy Phys.} {#1}, #3 (#2)}
\def\jphg#1 #2 #3{ {\rm J. Phys. G}           {#1}, #3 (#2)}
\def\cpc#1 #2 #3{ {\rm Comp. Phys. Comm.}    {#1}, #3 (#2)}  

\clearpage

\clearpage

\begin{figure}
\begin{center}
\hskip-5.0cm\epsfig{file=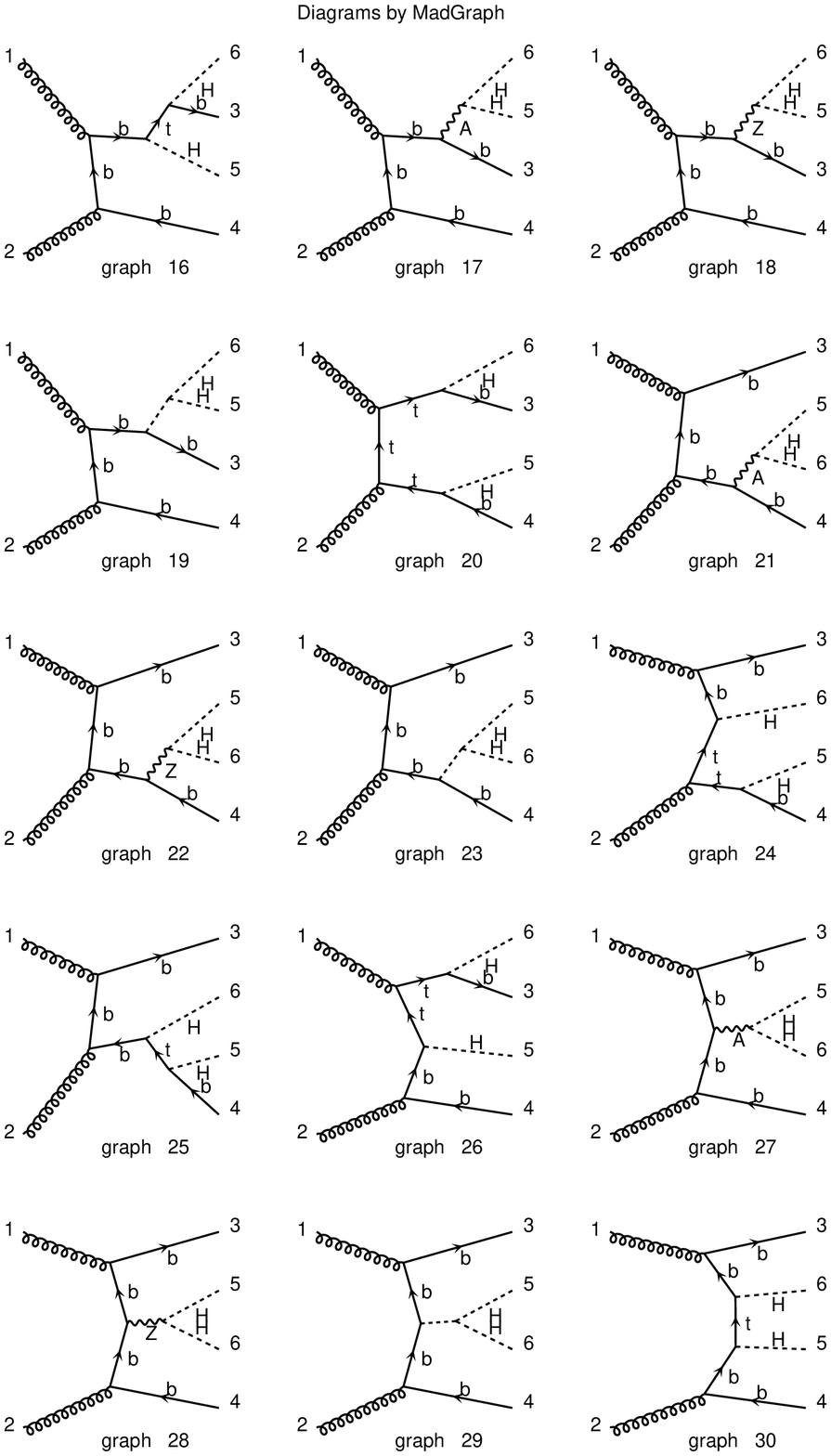,width=14cm,angle=0}
\end{center}
\vspace{-1.85cm}
\hskip2.5cm{Figure 1: continued}
\end{figure}

\clearpage

\begin{figure}
\begin{center}
\hskip-5.0cm\epsfig{file=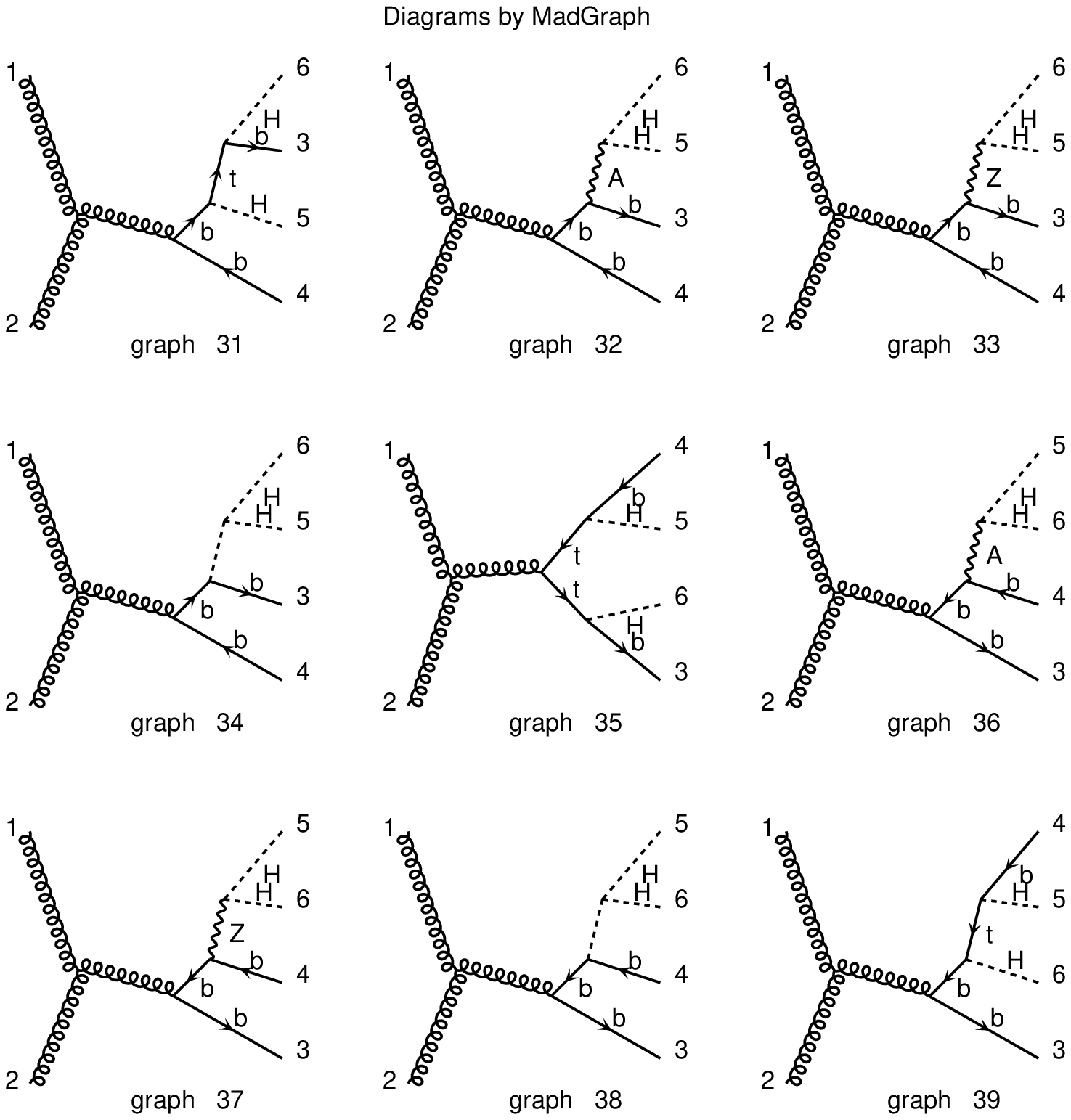,width=14cm,angle=0}
\end{center}
\vspace{-10.75cm}
\hskip2.5cm{Figure 1: continued}
\end{figure}

\clearpage
\begin{figure}
\begin{center}
\epsfig{file=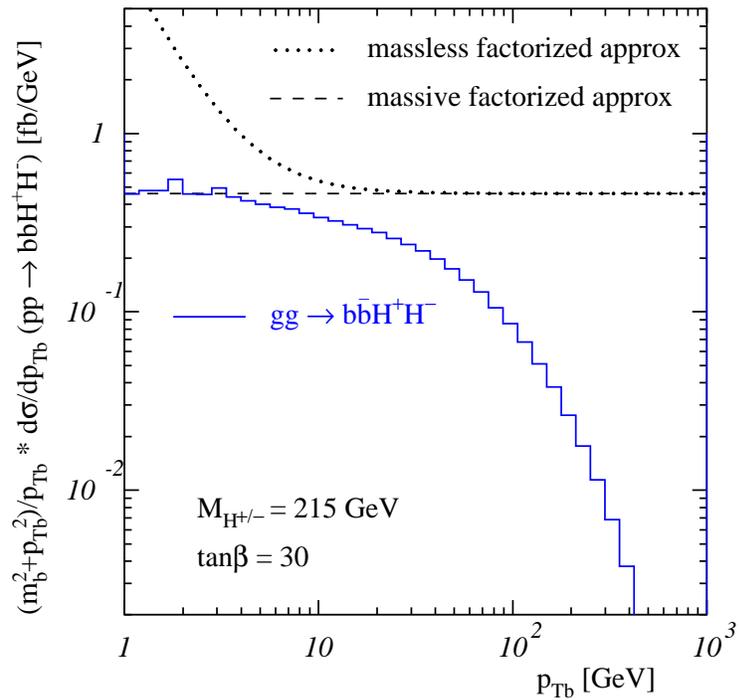,width=10cm,angle=0}
\end{center}
\caption{Transverse momentum distribution (multiplied by 
$((p^T_{b})^2+m_b^2)/p^T_{b}$) of $b$-quarks ($m_b=$4.25 GeV) in 
$gg \to b\bar{b}H^+H^-$ compared to factorised 
expectations for massless and massive partons.}
\label{fig:fact}
\end{figure}

\clearpage

\begin{figure}
\begin{center}
\epsfig{file=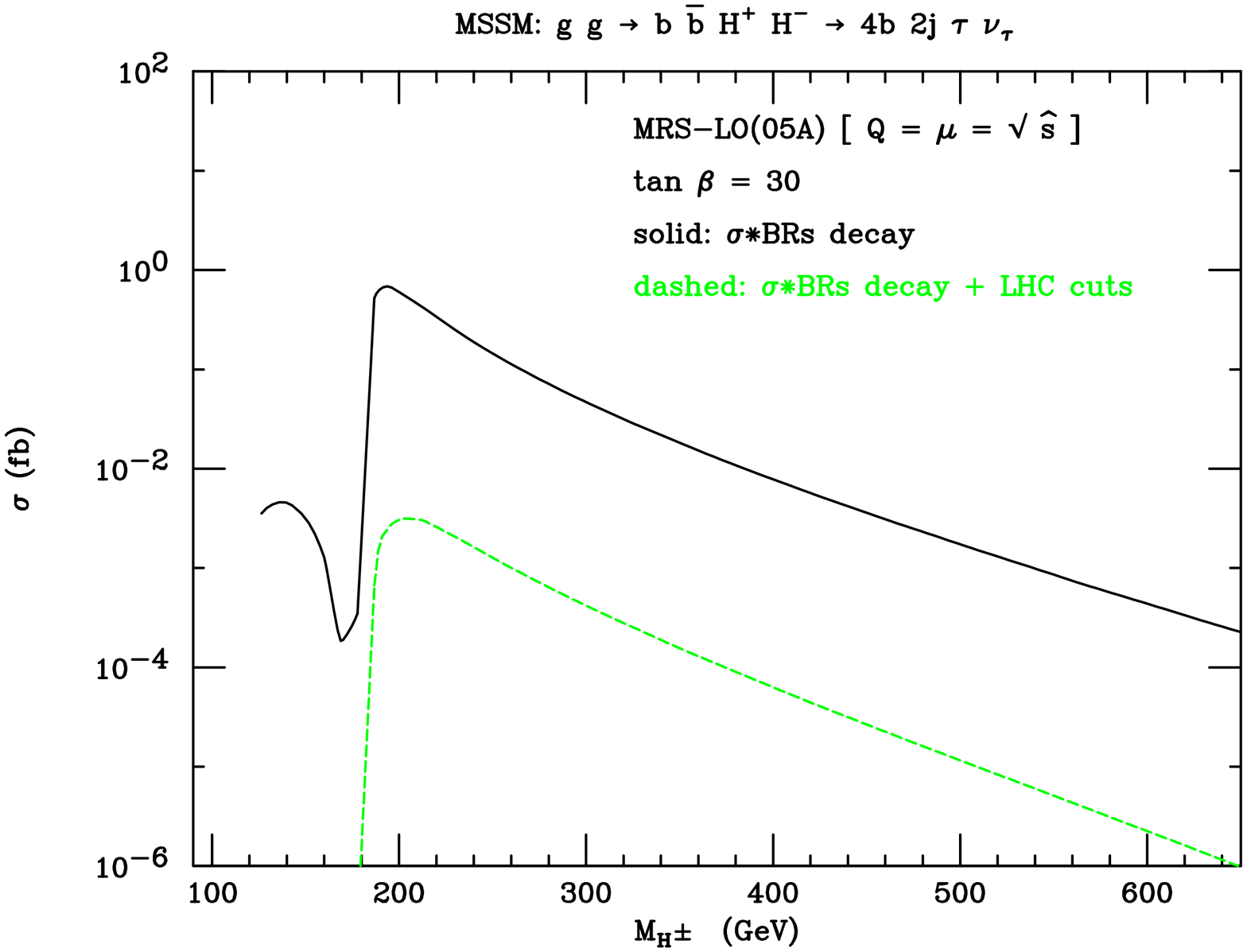,width=9cm,angle=90}
\end{center}
\vspace{-0.5cm}
\caption{Total cross sections as a function of $M_{H^\pm}$ for process 
(\ref{signal}) yielding the signature (\ref{signature}),
including  all decay BRs and with finite width effects, 
before (solid) and after the kinematic cuts in 
(\ref{cuts1})--(\ref{cuts5}), assuming LHC (dashed) 
detectors.
For reference, the value $\tan\beta=30$ is adopted. 
The MSSM described in the text is here assumed.}
\label{fig:bbHH_cuts}
\end{figure}

\clearpage

\begin{figure}
\begin{center}
\epsfig{file=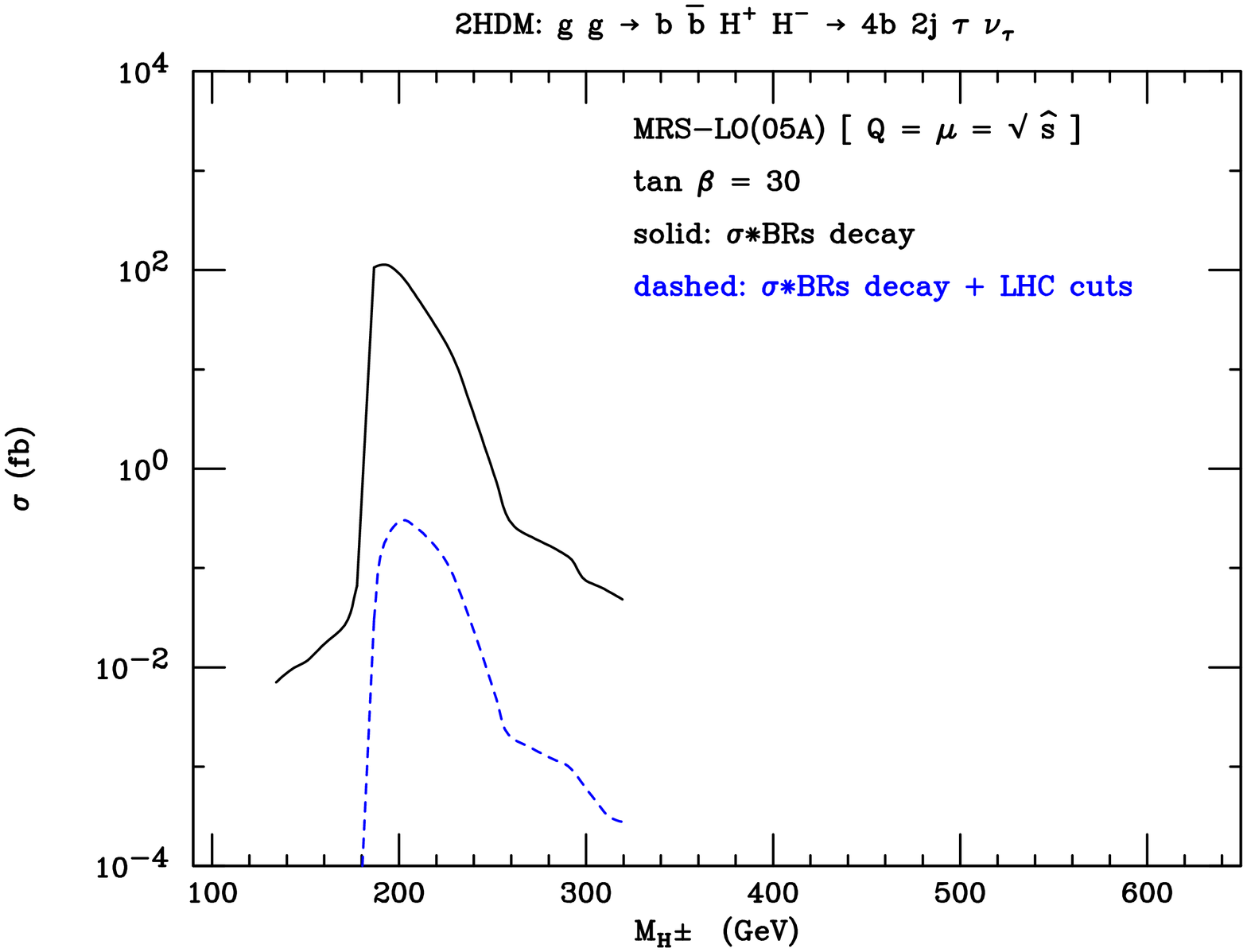,width=9cm,angle=90}
\end{center}
\vspace{-0.5cm}
\caption{Total cross sections as a function of $M_{H^\pm}$ for process 
(\ref{signal}) yielding the signature (\ref{signature}),
including  all decay BRs and with finite
width effects,
before (solid) and after the kinematic cuts in 
(\ref{cuts1})--(\ref{cuts5}), assuming LHC (dashed) 
detectors.
For reference, the value $\tan\beta=30$ is adopted. 
The 2HDM described in the text is here assumed.}
\label{fig:newbbHH_cuts}
\end{figure}

\clearpage

\begin{figure}
\begin{center}
\epsfig{file=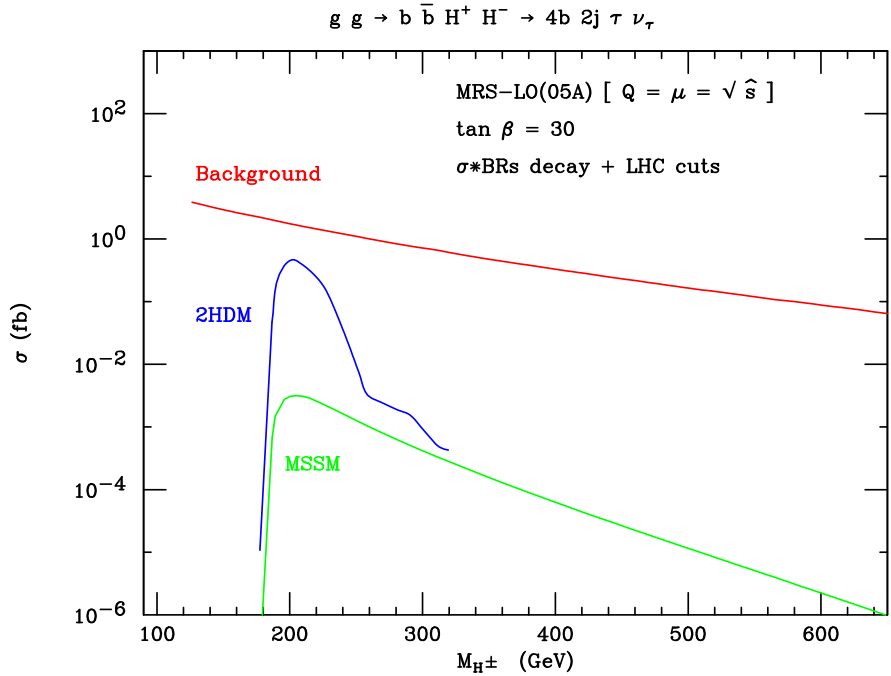,width=9cm,angle=90}
\end{center}
\vspace{-0.5cm}
\caption{Total cross sections as a function of $M_{H^\pm}$ for processes 
(\ref{signal}) (with finite width effects)
and (\ref{background}) yielding the signature (\ref{signature}),
including  all decay BRs and after the kinematic cuts in 
(\ref{cuts1})--(\ref{cuts5}), assuming LHC 
detectors.
For reference, the value $\tan\beta=30$ is adopted. 
The MSSM and the 2HDM described in the text are here compared 
to the background. (Notice that
the background retains a dependence upon $M_{H^\pm}$ because of
the optimisation of the cut in (\ref{cuts2}).)}
\label{fig:compare_cuts}
\end{figure}

\clearpage

\begin{figure}
\begin{center}
\epsfig{file=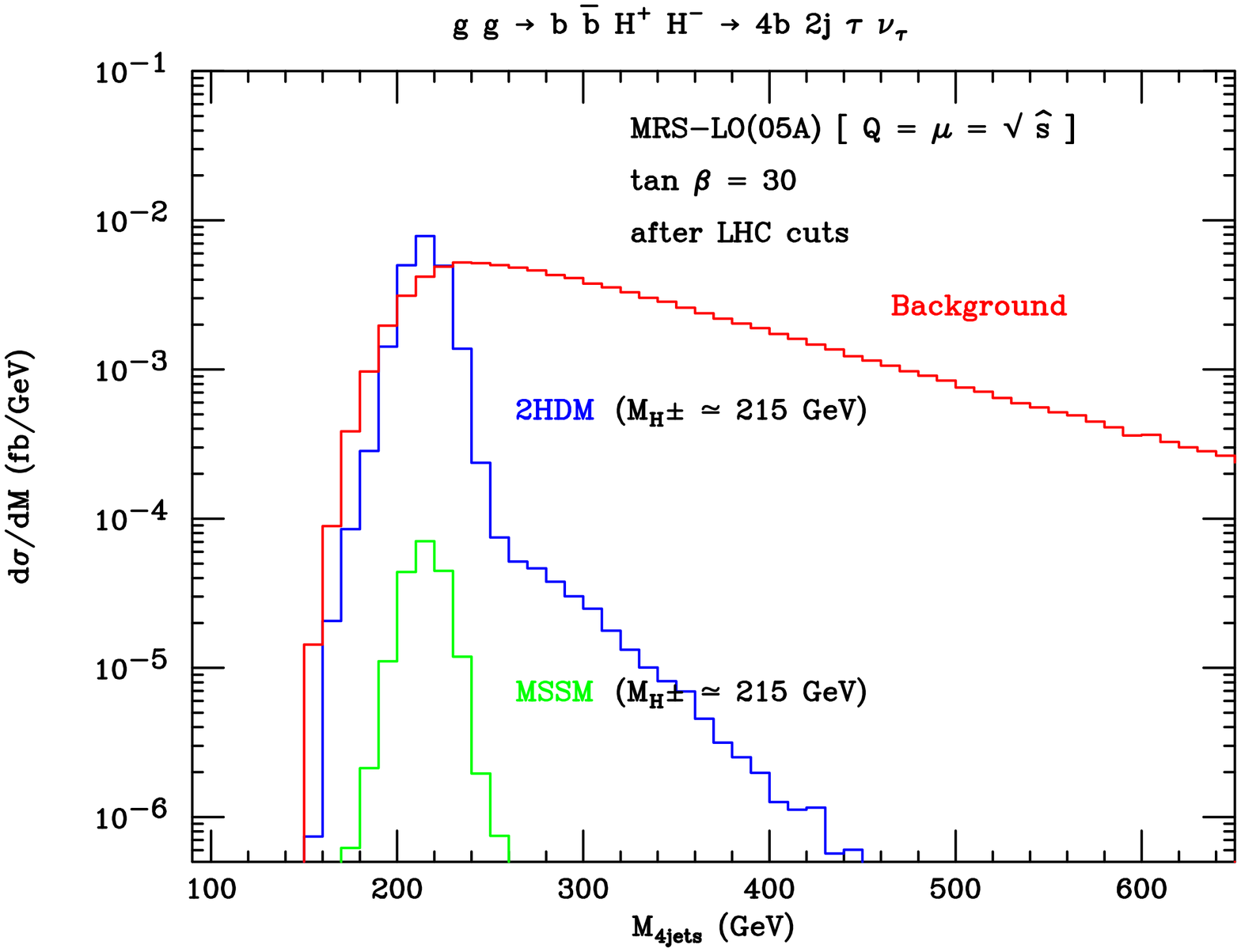,width=9cm,angle=90}
\end{center}
\vspace{-0.5cm}
\caption{Differential distribution in the invariant mass of the 
`four-jet system' defined in the text  for processes 
(\ref{signal}) (with finite width effects)
and (\ref{background}) yielding the signature (\ref{signature}),
including  all decay BRs and after the kinematic cuts in 
(\ref{cuts1})--(\ref{cuts5}), assuming LHC 
detectors.
For reference, the values $M_{H^\pm}=215$ GeV and $\tan\beta=30$ are adopted. 
The MSSM and the 2HDM described in the text are here compared to
the background.}
\label{fig:Hmass_cuts}
\end{figure}

\clearpage

\begin{figure}
\begin{center}
\epsfig{file=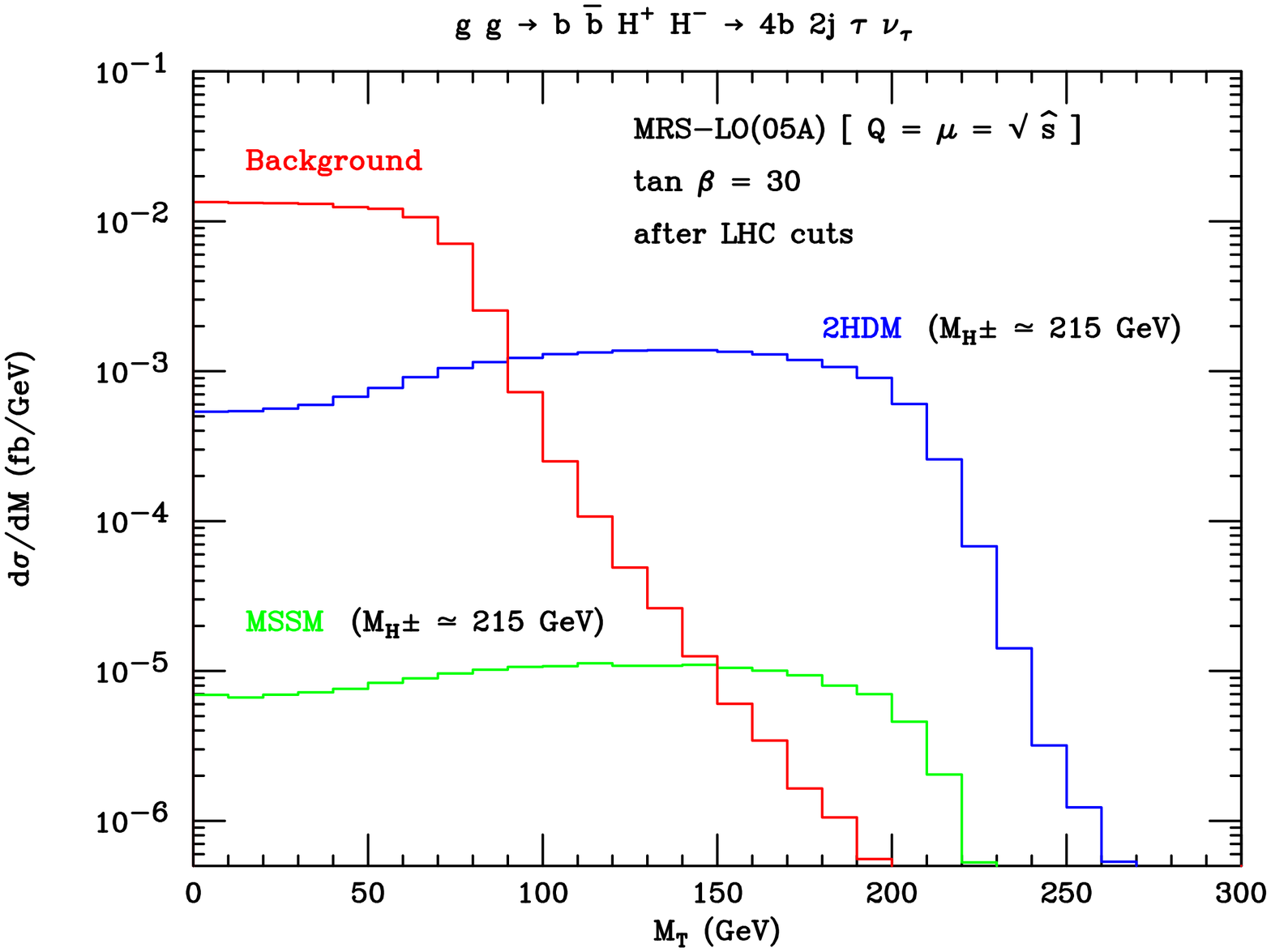,width=9cm,angle=90}
\end{center}
\vspace{-0.5cm}
\caption{Differential distribution in transverse mass of the `tau-neutrino'
system defined in the text for processes 
(\ref{signal}) (with finite width effects)
and (\ref{background}) yielding the signature (\ref{signature}),
including  all decay BRs and after the kinematic cuts in 
(\ref{cuts1})--(\ref{cuts5}), assuming LHC 
detectors.
For reference, the values $M_{H^\pm}=215$ GeV and $\tan\beta=30$ are adopted. 
The MSSM and the 2HDM described in the text are here compared to
the background.}
\label{fig:Tmass_cuts}
\end{figure}

\clearpage

\begin{figure}
\begin{center}
\epsfig{file=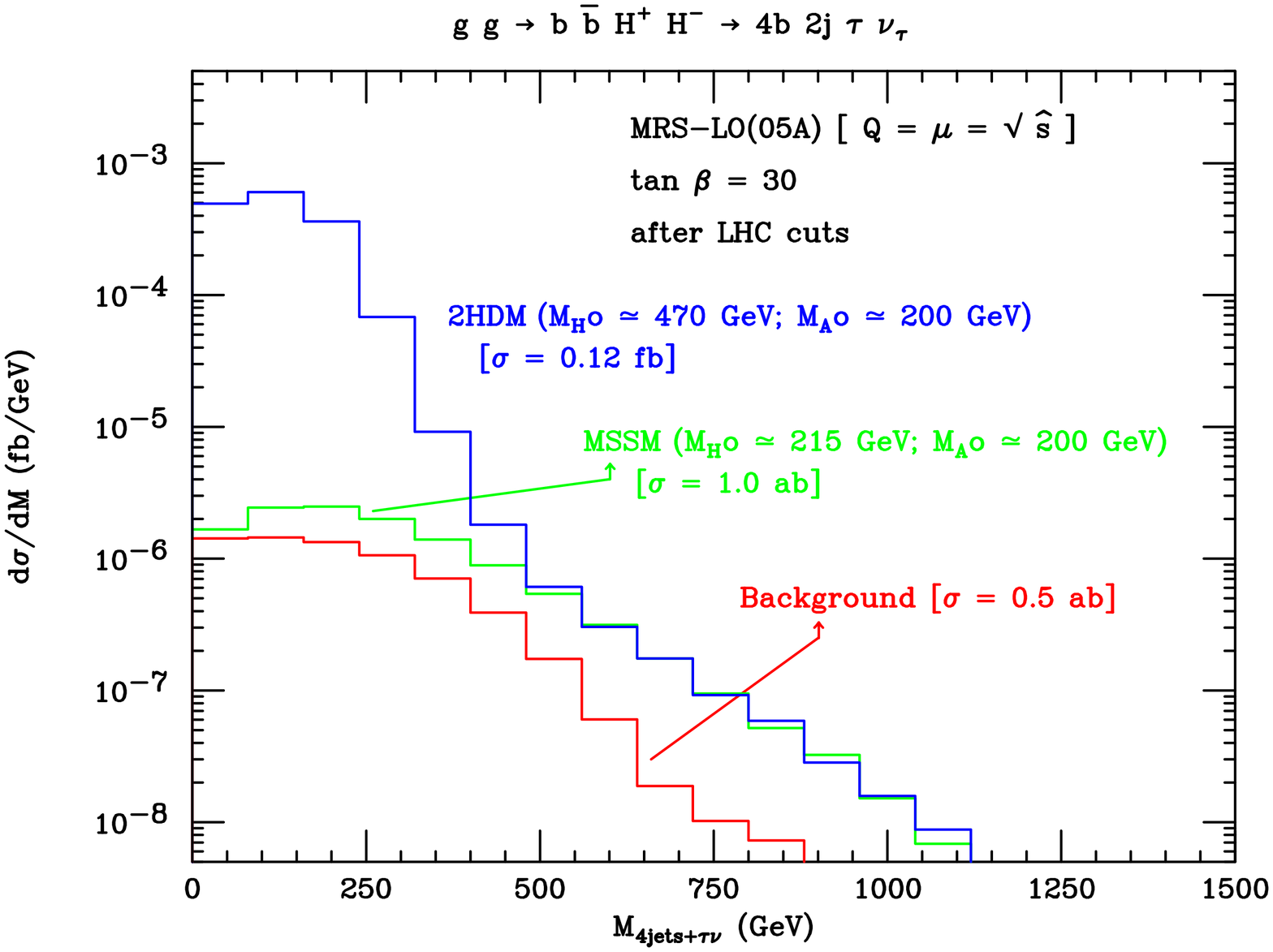,width=9cm,angle=90}
\end{center}
\vspace{-0.5cm}
\caption{Differential distribution in transverse mass of the 
`four-jets plus tau-neutrino'
system defined in the text for processes 
(\ref{signal}) (with finite width effects)
and (\ref{background}) yielding the signature (\ref{signature}),
including  all decay BRs and after the kinematic cuts in 
(\ref{cuts1})--(\ref{M4cut}) and (\ref{MTcut}), 
assuming LHC 
detectors.
For reference, the values $M_{H^\pm}=215$ GeV and $\tan\beta=30$ are adopted. 
The MSSM and the 2HDM described in the text are here compared to
the background.}
\label{fig:HHmass_cuts}
\end{figure}

\end{document}